\documentclass[journal]{IEEEtran}
\usepackage{graphicx}
\usepackage{amsmath}
\usepackage{booktabs}
\usepackage{tabularx}
\usepackage{float}
\usepackage{url}
\usepackage{placeins} 
\usepackage{underscore}

\begin{document}

\title{LK\_Jam: System Architecture and Implementation of a Real-Time Human-AI Interactive Music Generation System using Role-Aware GRU}

\author{Yakun~Liu, Z.~Jin, Dong~Liu, and~Hai~Luan
\thanks{Yakun Liu, Dong Liu, and Hai Luan are with the Shenyang Conservatory of Music. Z. Jin is an independent collaborator. (Corresponding author: Dong Liu.)}}

\maketitle

\begin{abstract}
As artificial intelligence advances into the era of Embodied AI, live musical interaction urgently needs to break free from the limitations of offline, unidirectional generation, achieving a "virtual synergy" capable of low-latency, dynamic interplay. To address this, this technical report presents LK\_Jam, a real-time, bidirectional human-computer interactive music generation system based on a lightweight Gated Recurrent Unit (GRU) and a high-performance audio host architecture. In the algorithmic representation layer, this system abandons the computationally expensive fixed time-grid. Instead, it constructs a multi-dimensional sparse event stream integrating time-shifts, continuous harmonic embeddings, and role-aware encoding, enabling the model to accurately capture turn-taking logic and micro-timing in a single-step inference. In the engineering implementation layer, this paper builds a strict multithreaded lock-free communication bridge using C++ and the JUCE framework, incorporating the RTNeural inference engine designed specifically for real-time audio. By utilizing compile-time network topology solidification and a zero-allocation (allocation-free) mechanism, the end-to-end overhead of autoregressive decoding is strictly locked at $O(1)$ complexity, structurally mitigating the risk of audio thread dropouts in DAW plugin environments. Furthermore, this study designs a three-stage progressive training strategy, achieving a leap from basic chord harmonization to expert-level interaction. Preliminary observations and architectural analysis demonstrate that while ensuring musical coherence and interactive role-play, the proposed system successfully challenges extreme real-time engineering constraints, offering a highly robust and deployable technical paradigm for next-generation AI co-performers in live music.
\end{abstract}


\section{Introduction}

\subsection{Background and Significance}
As artificial intelligence comprehensively shifts towards "Embodied AI" and deep learning technologies penetrate vertical application domains, the core dimension of interaction is driving the ultimate evolutionary form of music AI from offline "auxiliary generation tools" to "embodied co-performers" capable of playing alongside humans. To achieve such profound physical or virtual synergy, models must possess ultra-low-latency real-time response capabilities and complex interactive cognitive logic. However, despite significant breakthroughs in sequence coherence for symbolic music generation driven by deep learning, bridging the gap to true live interactive environments remains a massive challenge. Music is not merely a static accumulation of notes; its core vitality stems from immediate auditory feedback and dynamic interplay between musicians. In real live performances, the tacit cooperation based on the alternating balance of "listening" and "responding" directly determines the final output. This characteristic constitutes the core pain point of current interactive music AI research: how to enable models to transcend unidirectional generation, achieve accurate "role awareness," and satisfy the rigorous low-latency requirements of bidirectional interaction. Jazz improvisation, as a typical paradigm of such dynamic interaction, provides a rigorous reference framework for establishing scientific turn-taking logic and a solid material foundation for building high-quality interactive datasets \cite{dinverno2020}.

\subsection{Limitations of Existing Methods}
(1) \textbf{Algorithmic Computation Bottlenecks:} Existing mainstream generative models suffer from high inference latency in offline tasks due to their massive parameter counts, failing to meet the strict "low latency" and "real-time response" requirements of live improvisation. 
(2) \textbf{Lack of Interaction Logic:} Most interactive music AIs lack a distinct "role consciousness," clear division of labor, and turn-taking mechanisms. The generated phrases are often unidirectional continuations rather than genuine "dialogues" \cite{mccormack2024}. 
(3) \textbf{Engineering Deployment Barriers:} Traditional cross-disciplinary solutions face significant real-time obstacles when deployed in Digital Audio Workstations (DAWs). Embedding large, general-purpose inference engines directly into VST3 plugins introduces implicit multithreaded scheduling that easily preempts CPU resources, causing audio thread dropouts. Conversely, inter-process connection schemes relying on non-real-time communication protocols (e.g., OSC) inevitably introduce unacceptable system-level communication latency.

\subsection{Core Contributions}
To bridge the cognitive gap in deep human-computer interaction and the engineering barriers of edge audio inference, this report proposes a novel real-time interactive music AI architecture. The main contributions are:
\begin{itemize}
    \item \textbf{A 3D data encoding strategy featuring Role/Phrase Identifiers}, enhancing the temporal model's context-awareness in multi-turn interactions.
    \item \textbf{A low-latency audio engineering framework based on C++, JUCE, and RTNeural}, bridging the full deployment pipeline from offline AI model training to DAW real-time edge inference.
    \item \textbf{A three-stage progressive training strategy} based on algorithmic generation and expert-level alignment, aligning the AI's generative logic closer to the authentic improvisational mindset of human musicians.
\end{itemize}

This study integrates data feature representation with model architecture design, culminating in the engineering development of a JUCE-based audio plugin. The source code has been open-sourced and is publicly available on GitHub (\url{https://github.com/yakunliu-aimusic/LK_Jam}). As the project remains in an active iteration phase, this article serves as a preliminary technical report intended for academic demonstration and community exchange.

\begin{figure}[htbp]
    \centering
    \includegraphics[width=\linewidth]{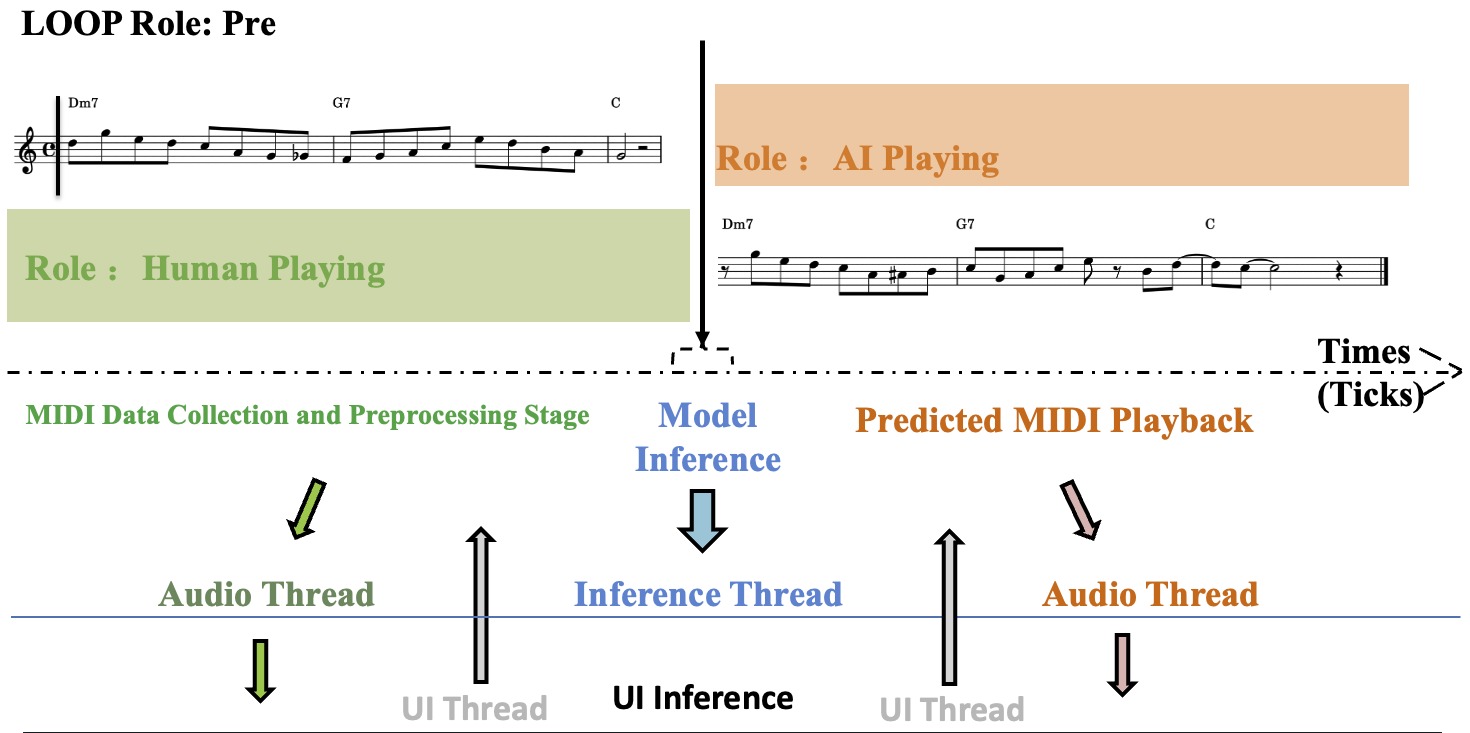} 
    \par\vspace{0.2cm}
    \includegraphics[width=\linewidth]{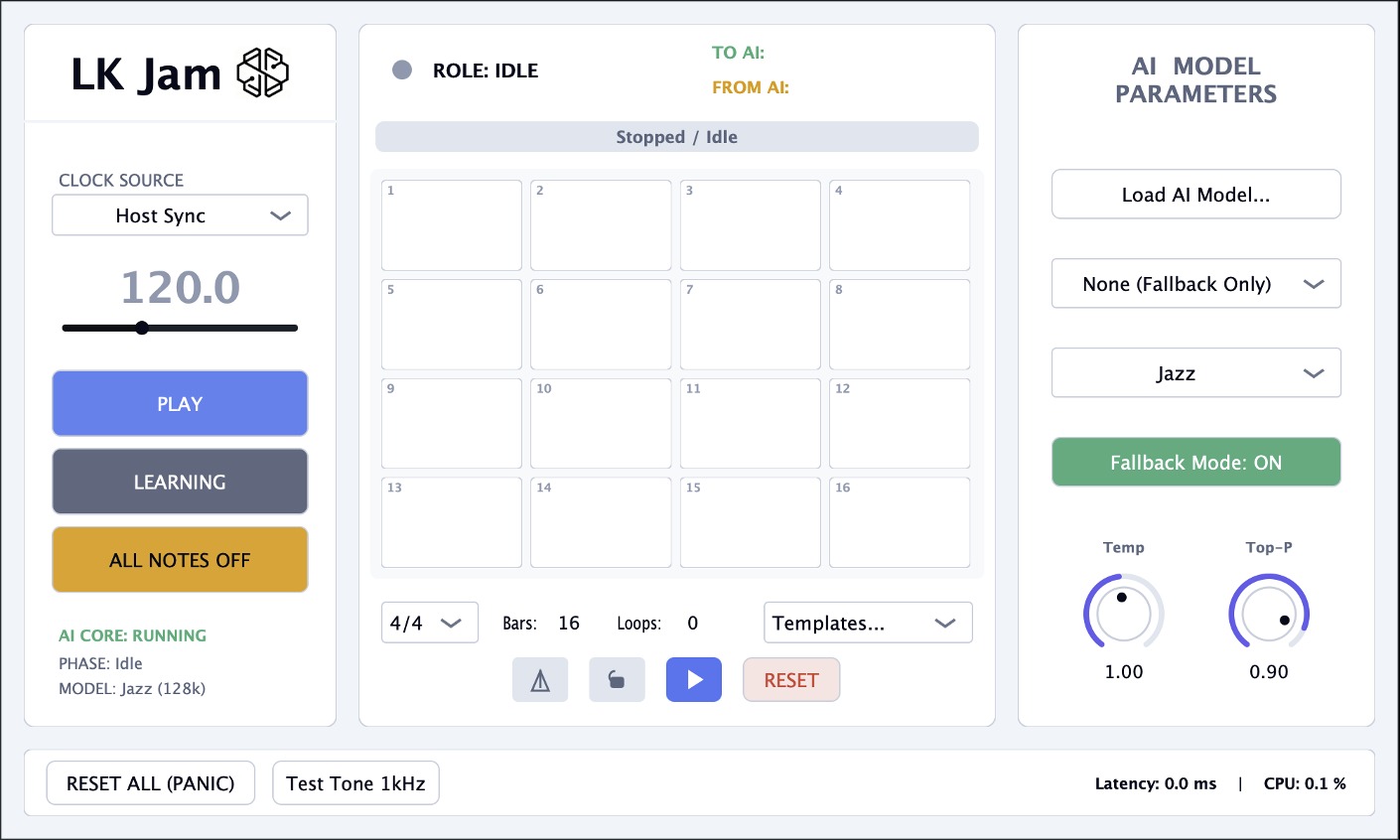}
    \caption{LK\_Jam System UI Interface and Interactive Architecture.}
    \label{fig:ui}
\end{figure}

\section{Related Work}

\subsection{Latency Bottlenecks and Architecture Evolution}
When exploring the underlying algorithms of automated composition systems, the inherent contradiction between model time complexity and the absolute real-time nature required for live improvisation is paramount. Early traditional algorithms (e.g., Markov Chains, HMMs) \cite{pachet2003} easily achieved zero-latency responses due to extremely low computational overhead. However, these algorithms are essentially shallow simulations of discrete state transition probabilities, lacking deep musical feature extraction and long-range semantic learning.

With the introduction of deep learning, symbolic music generation achieved qualitative leaps, but severe latency bottlenecks emerged. The currently dominant self-attention architectures (Transformer/GPT models) \cite{vaswani2017} excel in global structure generation. Still, their requirement to compute probability distributions over the entire sequence and vocabulary at every inference step results in an $O(N^2)$ computational complexity, rendering them unviable for direct application within the real-time audio callbacks of DAW plugins \cite{holtzman2019}.

In comparing sequence learning with inference speed, Temporal Convolutional Networks (TCN) \cite{bai2018} briefly became popular due to the large receptive fields provided by dilated convolutions and parallel computation. Yet, TCNs face an irreconcilable conflict between temporal resolution (TPQN) and computational load. Capturing the essence of musical micro-timing (e.g., Swing, Lay back) requires high temporal resolution rather than coarse, fixed sixteenth-note grids. Increasing TPQN for fine-grained expression causes the TCN's computational load to surge with sequence length, neutralizing its speed advantage.

Within the Recurrent Neural Network (RNN) family, while basic RNNs offer ultra-fast single-step responses, they suffer from vanishing gradients in long sequences. Long Short-Term Memory (LSTM) \cite{hochreiter1997} networks solve long-range dependencies, but their massive matrix operations remain too heavy for CPU-sensitive VST3 embedded audio threads. In contrast, the Gated Recurrent Unit (GRU) \cite{cho2014} retains memory capabilities while streamlining internal structures. Its single-step $O(1)$ time complexity makes it an optimal compromise for processing timestamp-based, high-resolution sparse event streams, providing the algorithmic prerequisite for subsequent lock-free acceleration via RTNeural.

\subsection{Role Blind Spots in Interactive Music}
Traditional interactive systems handling live engagement heavily rely on rule-based triggers or shallow mapping mechanisms. This interaction is essentially mechanical stimulus-response simulation; the system neither understands the "call and response" of musical motifs nor comprehends role division.

In the deep learning domain, existing interactive music AIs often become tireless "infinite continuers" \cite{mccormack2019}. This is not because temporal models lack the capacity to learn roles, but because existing research generally fails to adapt the underlying data structure for the vertical application of "live human-computer interaction." Current methods treat pieces as homogeneous symbol streams extending in a single direction, failing to explicitly encode "Role" and "Turn-taking" into the model's feature learning.

In broad live interaction paradigms, musicians constantly toggle between "listening" and "playing" states. While this system uses jazz "Trading Fours"—which demands rigorous interactive rules—as the testbed, this underlying turn-taking logic is highly generalizable to broader human-AI collaborative scenarios. Without prior data structures explicitly defining roles, existing models cannot determine their current interactive state, resulting in aimless phrase generation devoid of clear musical phrasing. Therefore, introducing role identifiers during the data representation stage is crucial to teaching the model when to advance and when to yield.

\subsection{Audio Host Plugins and Edge Inference}
In cross-disciplinary research, translating algorithmic models into DAW hosts faces an enormous real-time chasm. Early integration schemes relied heavily on external non-real-time protocols (e.g., OSC or virtual MIDI ports) to bridge the host with isolated Python inference processes. These schemes, completely dependent on the OS's general thread scheduling, cannot withstand uncontrollable jitter caused by system load fluctuations, introducing massive inter-process latency entirely unsuitable for rapid jazz improvisation.

Recent attempts (e.g., general AI audio plugin frameworks like Neutone) aim to encapsulate general deep learning engines (like ONNX Runtime or TensorFlow Lite) directly within VST3/AU plugins. However, as Yee-King \cite{yeeking2024} points out regarding AI-enhanced audio plugin architectures, a DAW's `processBlock` operates on a high-priority real-time thread with an extremely tight time budget. General engines like ONNX are not only bulky but also implicitly contain uncontrollable thread pools. Furthermore, they may trigger Dynamic Memory Allocation during forward passes. Such non-deterministic runtime behaviors frequently cause lock contention, preempting CPU cores and directly leading to audio dropouts \cite{yeeking2024}.

To address this edge inference pain point, this architecture completely abandons heavy general engines, pivoting to RTNeural \cite{chowdhury2021}, a lightweight C++ inference library designed specifically for real-time audio thread safety. Its core advantage lies in using C++ template metaprogramming to solidify network topologies and weights at compile-time. This grants the model true zero-allocation and wait-free execution capabilities at runtime, eradicating the risk of thread blocks caused by memory fragmentation or implicit scheduling, providing the optimal host inference environment for our lightweight GRU.

\section{Role-Aware Model Design}

\subsection{Closed-Loop Turn-Taking Logic}
The system draws inspiration from the turn-taking and call-and-response concepts of jazz, generalizing traditional Trading Fours into configurable loop-level human/AI alternations. Each complete loop is treated as a turn unit: even loops are played by humans, and odd loops are responded to by the AI. The workflow relies on a closed-loop bidirectional linkage: the human inputs a sequence $X_{human}$ of length $L$, and the model infers a response sequence $Y_{ai}$ under a given harmonic condition $C$. The human then bases their next input on $Y_{ai}$, forming a continuous musical dialogue.

Given the monophonic linear nature of live jazz solos, the system compresses multi-dimensional information into a single-track event stream rather than employing complex multi-voice representations. Bounding the improvisational nature this way allows the model to shed redundant calculations, ensuring the $O(1)$ complexity required for zero-latency response on an embedded audio thread.

\subsection{Timestamp Sparse Event Streams \& 3D Features}
When processing jazz improvisation with high degrees of freedom and strong micro-timing (e.g., swing, lay back), forced quantization alignment using fixed 16th- or 32nd-note grids erases the human performance's groove and generates massive amounts of zero-padding due to high Time Pulses Per Quarter Note (TPQN). This data sparsity explosion consumes immense computational power and impedes audio thread inference efficiency.

Consequently, the system completely abandons fixed grids in favor of an event-triggered Time-shift ($\Delta t$) sparse event stream format \cite{oore2020}. Each valid MIDI note is modeled as an independent time step $t$, and the model only triggers computation upon receiving an actual note event.

To enable the GRU to simultaneously perceive melody, dynamics, timing, harmony, and interaction logic in a single inference step, the system constructs a multi-dimensional composite input feature vector $x_t$. At time step $t$, the input tensor is formed by concatenating several embedding modules:

\begin{equation}
\begin{aligned}
x_t = \text{Concat}\big( &E_{pitch}(p_t), E_{vel}(v_t), E_{time}(\Delta t_t), \\
                         &C_t, E_{role}(r_t), E_{phrase}(loc_t) \big)
\end{aligned}
\end{equation}

\begin{itemize}
\item $E_{pitch}(p_t)$: Dense vector mapping of the current Pitch.
\item $E_{vel}(v_t)$: Vector mapping of the note Velocity (0-127), enabling the model to learn accent distribution and dynamics.
\item $E_{time}(\Delta t_t)$: Time-shift embedding between the current and previous note onset, ensuring the model learns complex relative rhythmic relationships.
\item $C_t$: Current harmonic context, utilizing Continuous Harmonic Space Embedding \cite{tshe} to help the model transcend rigid chord matching and understand tension trajectories.
\item $E_{role}(r_t) \in \{0, 1\}$: Role-aware encoding. Hardcoded to identify whether the current input belongs to the Human or the AI.
\item $E_{phrase}(loc_t)$: Phrase position identifier (Start, Continue, End). Forces the model to learn the phrasing arc of human motifs, ensuring the AI outputs closed-loop phrases with clear cadences rather than endless note sprawl.
\end{itemize}

\subsection{Lightweight GRU Round-Robin Generation}
In the "listen-respond" interactive framework, the AI does not perform note-by-note real-time accompaniment during the human's performance. Instead, at the boundary where the human turn ends, it centrally preprocesses the collected MIDI event stream and rapidly infers the complete response sequence for the next turn.

To ensure fluid and stable turn-taking and avoid groove discontinuities caused by model computation overload, the system employs a lightweight GRU to drive core hidden state updates:
\begin{equation}
    h_t = \text{GRU}(x_t, h_{t-1})
\end{equation}
The core advantage of this design is that a single-step forward pass strictly relies only on the current environmental tensor $x_t$ and the condensed memory $h_{t-1}$. This stateful computation, locked at $O(1)$ complexity, completely avoids the surging computational costs associated with autoregressive models (like Transformers) as sequences lengthen. It enables the background inference thread to complete the decoding of an entire phrase "faster-than-realtime" within the extremely brief window of turn transition. 

Finally, the hidden state is projected into a probability distribution via linear mapping $P(y_t) = \text{Softmax}(W_o h_t + b_o)$ for sampling. This minimalist generation pipeline seamlessly accommodates sparse event streams and decouples inference from the high-priority audio playback \cite{cho2014}.

\section{Progressive Training Strategy}

\subsection{Dataset Construction and Micro-Design}
The dataset employs a hierarchical construction strategy, corresponding to a three-stage training process:
1. \textbf{Basic Layer:} Algorithmically batch-generated monophonic melodies strictly constrained by core harmonic progressions (e.g., 2-5-1).
2. \textbf{Advanced Layer:} Integration of public jazz solo datasets, expanding to diverse harmonic scenarios (1-6-2-5, Blues) and adding non-chord tones and syncopations.
3. \textbf{Expert Layer:} Human-computer interactive motif call-and-response materials composed by professionals, covering logic like imitation and inversion.
The entire dataset utilizes 12-key transposition for data augmentation, ensuring the model learns harmonic relationships rather than fixed pitches \cite{bengio2015}.

\subsection{Three-Stage Progressive Training Pipeline}
Scheduled Sampling is introduced throughout training to gradually reduce Teacher Forcing, mitigating exposure bias in autoregressive models and improving long-phrase structural integrity.

\subsubsection{Stage 1: Atomic Harmonization}
\begin{figure}[htbp]
    \centering
    \includegraphics[width=\linewidth]{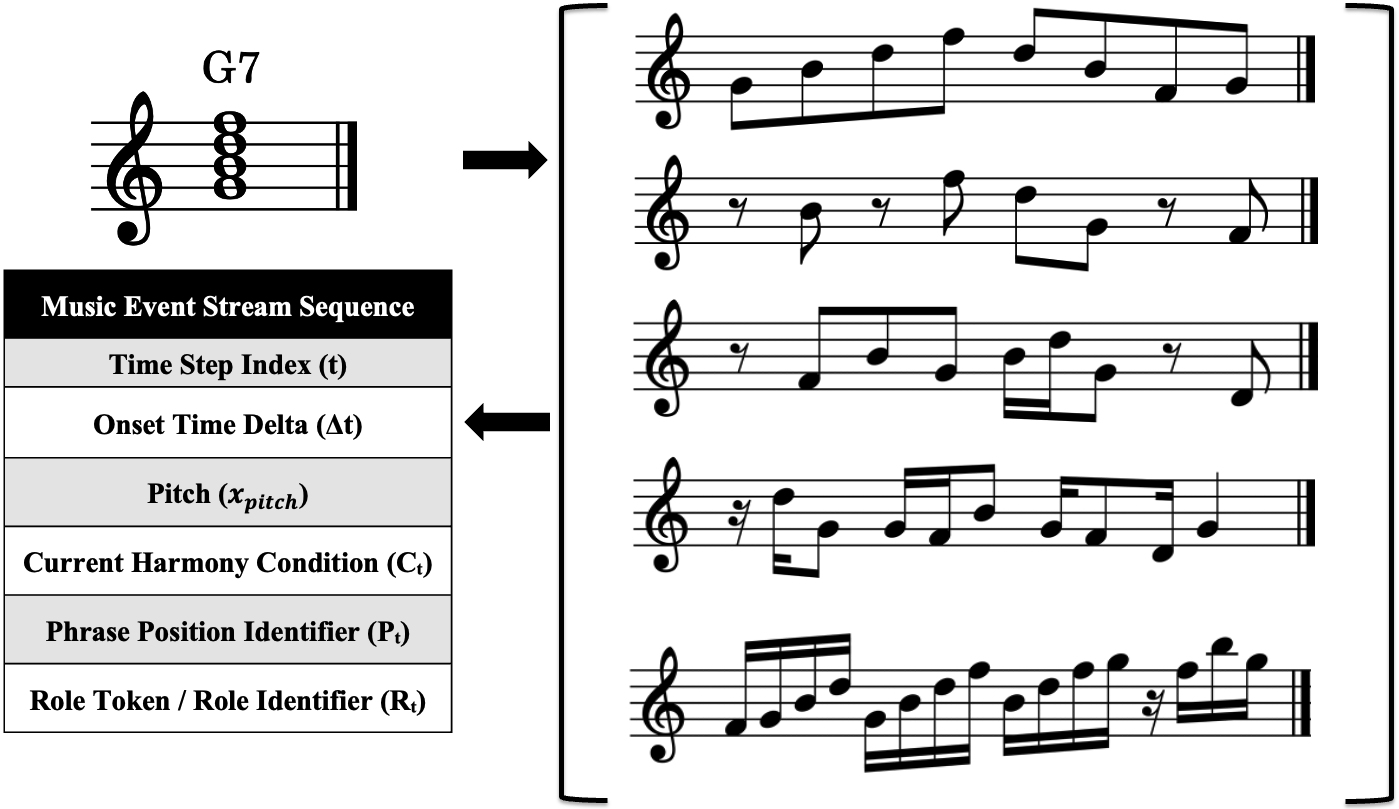}
    \caption{Stage 1: Atomic Harmonic Mapping Pre-training. Contains no non-chord tones or complex syncopation, purely enforcing the $C_t \rightarrow x_{pitch}$ atomic spatial mapping \cite{harmonization}.}
    \label{fig:stage1}
\end{figure}
Using chord markers as anchors, dense single-measure datasets are batch-generated to force strong binding relationships between chord roots, thirds, sevenths, and $x_{pitch}$. 

\subsubsection{Stage 2: Stylistic Wandering \& Linear Vocabulary}
\begin{figure}[htbp]
    \centering
    \includegraphics[width=\linewidth]{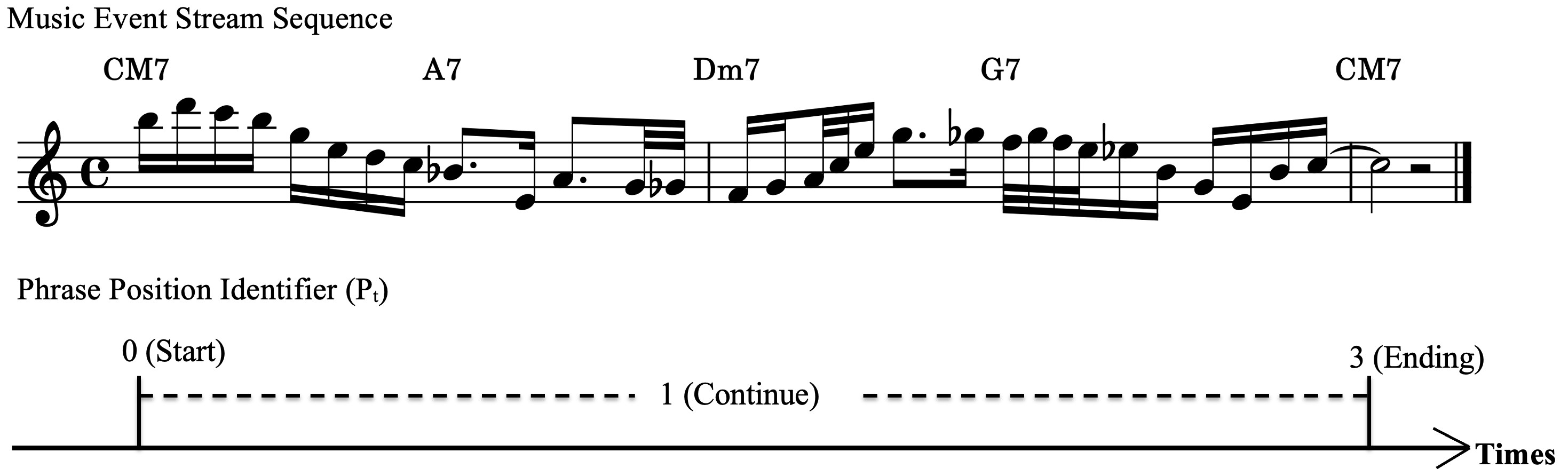}
    \caption{Stage 2: Linear Vocabulary Expansion and Phrase Self-Motif Development. Introduces chromatic approaches, Bebop scales, and syncopation to enhance syntactical logic.}
    \label{fig:stage2}
\end{figure}
Once the harmonic framework is stable, this stage introduces mainstream harmonic materials (e.g., ii-V-I) and classic jazz solo corpora. Target techniques such as Approach Notes, Enclosures, and Bebop scales \cite{levine1995} are added to the feature learning dimensions. Furthermore, self-motif development techniques (e.g., sequencing, rhythmic displacement) within a single turn are introduced, granting the AI independent syntactical logic.

\subsubsection{Stage 3: Expert-Level Interaction Alignment}
\begin{figure}[htbp]
    \centering
    \includegraphics[width=\linewidth]{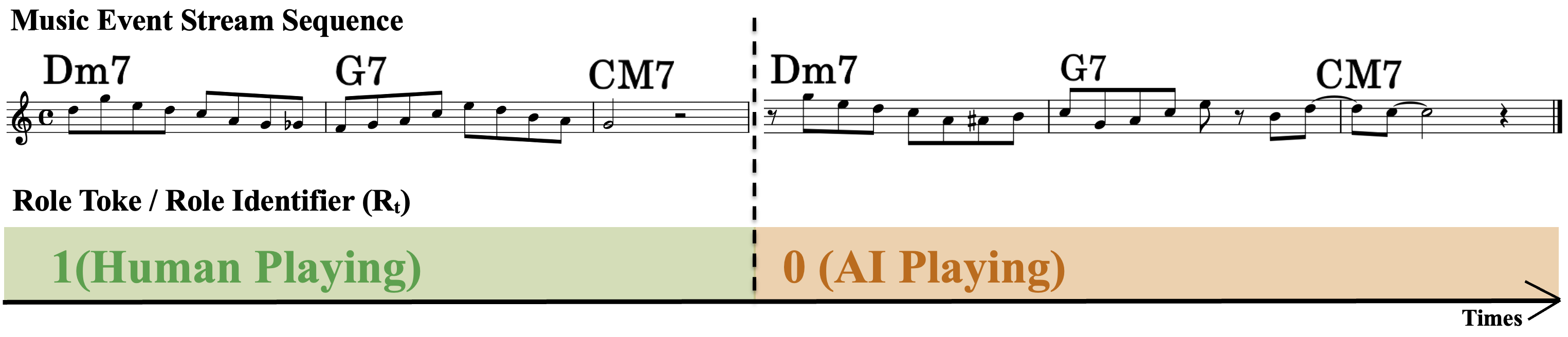}
    \caption{Stage 3: Expert-Prior Interactive Logic Alignment. Activates Role-Aware Encoding using Trading Fours corpora customized by domain experts, crossing from unidirectional generation to bidirectional interplay.}
    \label{fig:stage3}
\end{figure}
This final stage utilizes Supervised Fine-Tuning (SFT) with high-quality interactive materials to fully activate the "Role-aware Encoding" ($E_{role}$). It focuses on cross-turn bidirectional interplay. By strictly aligning and hardcoding motif response pairs like "Imitation," "Inversion," and "Motif Response" into the training sequences, the model is forced to deeply understand the "listen-analyze-respond" cycle. This thoroughly breaks the temporal model's tendency towards infinite sprawl, achieving a true closed-loop musical dialogue.

\section{Real-Time Engineering Architecture}

\subsection{JUCE-based Triple-Thread Model}
To meet real-time computational demands, the system is reconstructed using a low-level C++ architecture, discarding traditional non-real-time communication. Following JUCE framework best practices \cite{juce}, a strict classic triple-thread architecture is implemented (Audio Thread, Message/UI Thread, Worker/Inference Thread). The `processBlock` is completely decoupled from UI rendering. CPU-heavy AI generative tasks are securely placed in a background inference thread, ensuring the real-time audio thread only handles lightweight MIDI capture and playback scheduling.

\begin{table}[htbp]
\centering
\caption{Multithreaded Architecture Design}
\begin{tabularx}{\linewidth}{l X l}
\toprule
\textbf{Thread Type} & \textbf{Timestamp Purpose} & \textbf{Synchronization} \\
\midrule
Audio Thread & Real-time MIDI capture, state scheduling, playback. & Lock-free queue + Atomics \\
Inference Thread & Model forward pass, feature concatenation, inference. & Atomic double buffering \\
UI Thread & State machine rendering, UI lighting, note display. & 30Hz Timer Polling \\
\bottomrule
\end{tabularx}
\end{table}

\subsection{Bottom-Layer Lock-Free Communication Bridge}

\begin{figure*}[htbp]
    \centering
    \includegraphics[width=0.85\textwidth]{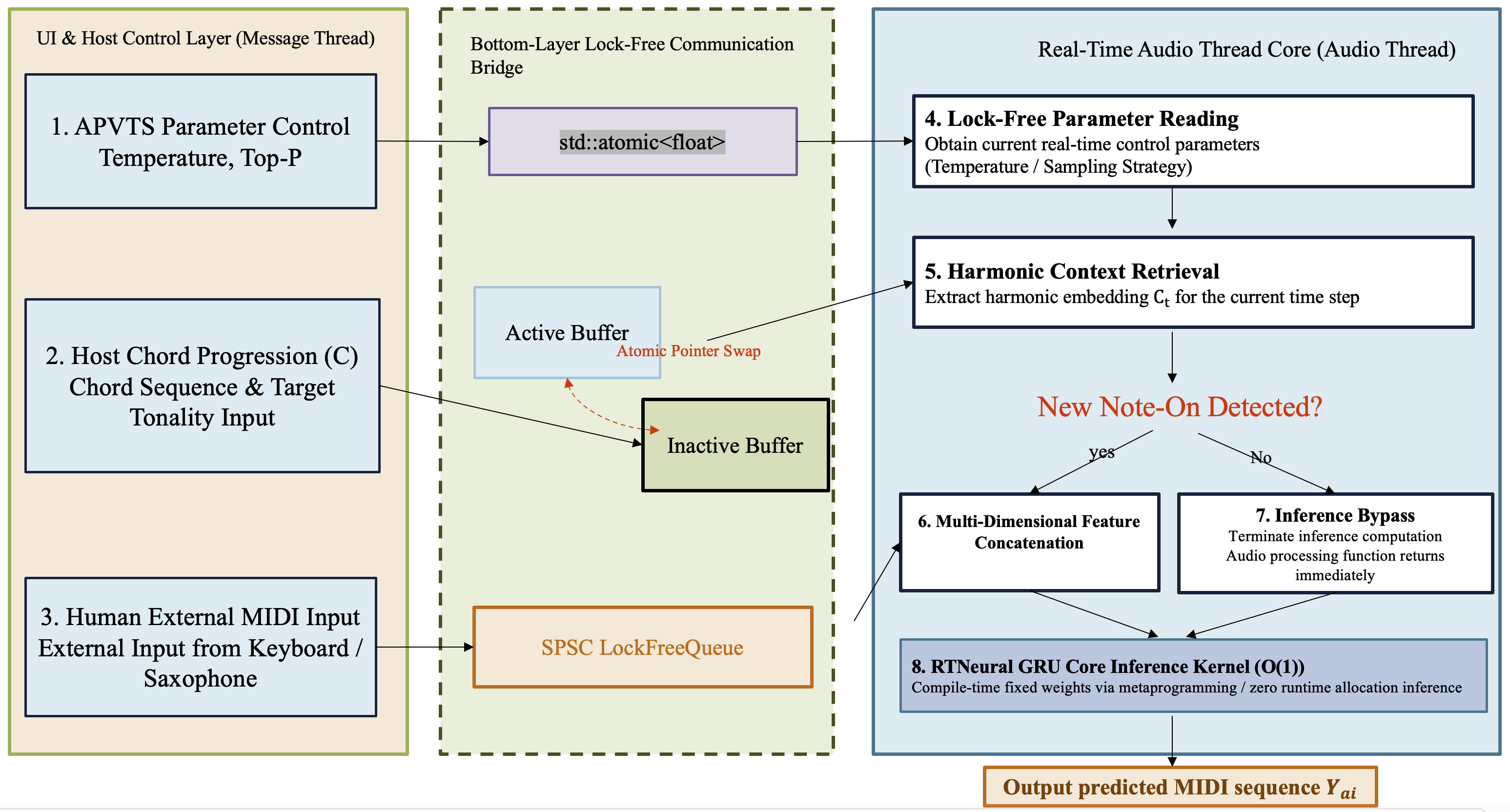}
    \caption{High-Performance Lock-Free Communication Architecture.}
    \label{fig:lockfree}
\end{figure*}

\subsubsection{Monitor Design (UI Optimization)}
To prevent the UI thread from consuming excessive system resources, the monitor employs a strategy of retaining a maximum of 5 note updates per cycle, lowering the graphics rendering overhead during 30Hz safe polling.

\subsubsection{Double Buffering for Heavy Weight Data}
For complex data structures like chord progressions, dual buffers are pre-allocated. The UI only writes data to the inactive buffer. Upon completion, an atomic pointer swap instantly provides the audio thread with the latest harmonic context without locking.

\subsection{Host PPQ Timeline Non-Blocking Sync}

\begin{figure*}[htbp]
    \centering
    \includegraphics[width=0.85\textwidth]{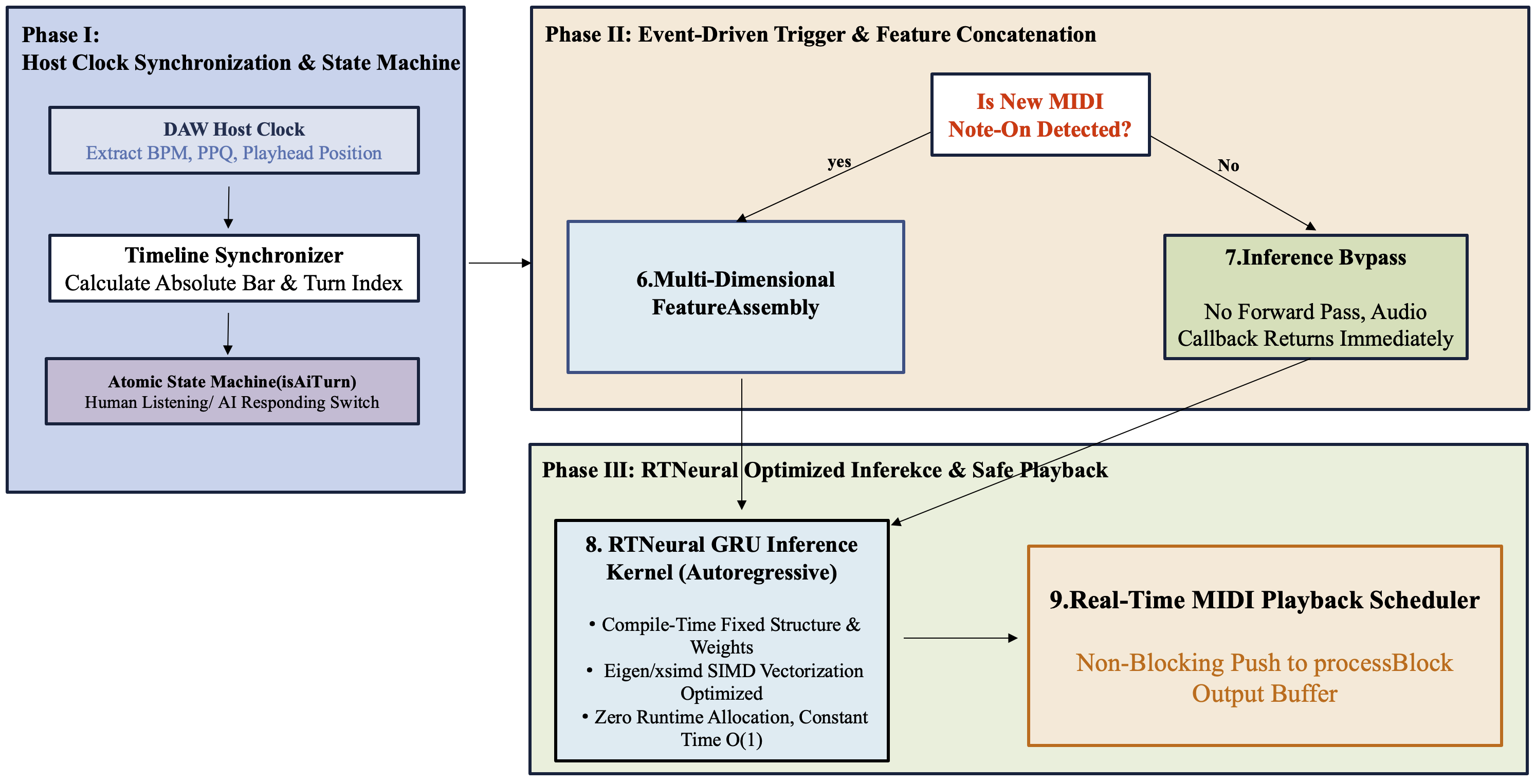}
    \caption{Real-Time Audio Thread Inference Pipeline.}
    \label{fig:pipeline}
\end{figure*}

By mounting onto the DAW's host clock, the synchronization engine extracts the current Pulses Per Quarter Note (PPQ) and BPM to calculate precise bar and turn indices. In host sync mode, the plugin automatically enters a PreRoll state one bar in advance to load harmonic contexts, seamlessly transitioning into the human turn once the relative PPQ hits zero at the start of the next formal loop. This ensures the Role-switching logic is strictly driven by the loop cycle and host timeline rather than an independent internal clock.

\subsection{RTNeural Compile-Time Solidification}
The system exclusively relies on RTNeural. Through C++ template metaprogramming, GRU weight dimensions and network topologies are solidified at compile-time. This achieves true allocation-free execution at runtime, effectively minimizing the risk of thread suspension caused by dynamic memory allocation during audio callbacks. Combined with vectorized instruction sets (Eigen/xsimd), it squeezes out underlying compute power to achieve absolute minimum latency \cite{chowdhury2021}.

\section{Expected Outcomes and Evaluation Plan}

\subsection{Architectural Performance Expectations}
Thanks to the dual-thread mechanism, lock-free SPSC queues, and RTNeural's compile-time solidification, the peak CPU usage of the audio thread is theoretically strictly constrained, even during full DAW playback. Compared to architectures relying on inter-process communication, this structural design drastically reduces the probability of thread blocks and audio dropouts caused by deep learning inference.

\subsection{Qualitative Evaluation and Planned Ablation Studies}
Following the proposed three-stage progressive training, the model's perplexity is expected to exhibit a stable downward trend \cite{ircam2023}. Future ablation studies are designed to verify whether introducing $E(R_t)$ (Role) and $E(P_t)$ (Phrase) significantly improves the thematic coherence and structural integrity of AI-generated phrases. Theoretically, stripping away the phrase identifier would predictably cause the generated melodies to infinitely extend, lacking breathability and definitive cadences.

\subsection{Subjective Auditory Evaluation Framework}
Referencing recognized collaborative criteria in international academia \cite{ircam2023}, this study establishes the six evaluation dimensions shown in Table \ref{tab:dimensions}, providing an assessment framework for future large-scale blind listening tests and A/B preference testing.

\begin{table*}[htbp]
\centering
\renewcommand{\arraystretch}{1.5}
\setlength{\tabcolsep}{14pt} 
\caption{Six Evaluation Dimensions for Human-Computer Improvisation Testing}
\label{tab:dimensions}
\begin{tabularx}{\linewidth}{c >{\raggedright\arraybackslash\hsize=0.8\hsize}X >{\raggedright\arraybackslash\hsize=1.2\hsize}X >{\raggedright\arraybackslash\hsize=1.0\hsize}X}
\toprule
\textbf{No.} & \textbf{Dimension Name} & \textbf{Core Definition} & \textbf{Model Training Goal} \\
\midrule
1 & Motif Response Strategy & The AI's interactive stance towards human motifs; the core logic of interplay. & Master motif development: Continuation, Imitation, Inversion, Conflict. \\
2 & Phrase Boundary & Standardizing start/end anchors to unify the symmetrical structure of phrasing. & Accurately identify $P_t$=Start and $P_t$=End to form closed loops. \\
3 & Harmonic Adherence & Measuring how well notes adhere to the current chord ($C_t$). & Distinguish chord tones, tensions, and altered notes to control tension. \\
4 & Linear Ornamentation & Collection of professional melodic refinement techniques. & Master passing tones, chromatic approaches, and double enclosures. \\
5 & Rhythmic Density & Controlling the forward momentum and dynamic pacing of the phrase. & Master rhythmic diminution/augmentation and syncopated displacement. \\
6 & Melodic Contour & Contrasting overall line shapes and register spans. & Retain core contours of human motifs without deviating from overall trajectory. \\
\bottomrule
\end{tabularx}
\end{table*}

\section{Discussion and Future Work}
This report demonstrates an architecture that bridges algorithmic design and engineering deployment for real-time interactive AI. To pursue extreme low latency and lightweight execution, the current architecture compromises on a monophonic event stream. Future work will explore lightweight multi-track concurrent networks (e.g., decoupled generation of chord and melody voicings) without compromising the lock-free nature of the VST3 real-time thread. Additionally, addressing the memory decay phenomenon in GRUs over extremely long structures, we plan to investigate the feasibility of running a slow "macro-structural planning model" in a non-real-time background thread to guide real-time generation.

\section{Conclusion}
The proposed AI improvisation plugin based on role awareness and a high-performance architecture validates that turn-taking mechanisms substantially enhance the interactivity of generated motifs. On an engineering level, it proves the massive potential of timestamp representations and lightweight deep learning models via the RTNeural engine in host audio streams, providing a highly robust technical paradigm for next-generation AI co-performers in live music.

\FloatBarrier 

\section*{Appendix: Interactive Data Representation Examples}

Example mapping across a complete C Major 7 (CM7) harmonic context.

\begin{figure}[htbp]
    \centering
    \includegraphics[width=\linewidth]{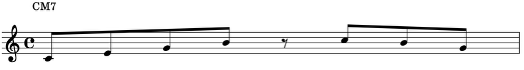}
    \caption{Example 1: Human Performance Motif Score}
    \label{fig:example1}
\end{figure}

\begin{table}[htbp]
\centering
\renewcommand{\arraystretch}{1.3} 
\caption{Human Playing Stage (Role: Human, corresponding to Example 1)}
\resizebox{\linewidth}{!}{
\begin{tabular}{l ccccccc}
\toprule
\textbf{Event} & \textbf{E 1} & \textbf{E 2} & \textbf{E 3} & \textbf{E 4} & \textbf{E 5} & \textbf{E 6} & \textbf{E 7} \\
\midrule
Time Step $t$ & $t_1$ & $t_2$ & $t_3$ & $t_4$ & $t_5$ & $t_6$ & $t_7$ \\
Time Delta $\Delta t$ & 0.0 & 0.5 & 0.5 & 0.5 & 1.0 & 0.5 & 0.5 \\
Pitch $x_{pitch}$ & 60 & 64 & 67 & 71 & 72 & 71 & 67 \\
Harmony $C_t$ & CM7 & CM7 & CM7 & CM7 & CM7 & CM7 & CM7 \\
Role $R_t$ & 1 & 1 & 1 & 1 & 1 & 1 & 1 \\
Position $P_t$ & Start & Continue & Continue & Continue & Continue & Continue & End \\
\bottomrule
\end{tabular}}
\end{table}

\begin{figure}[htbp]
    \centering
    \includegraphics[width=\linewidth]{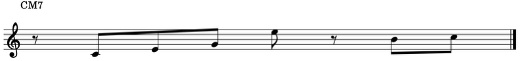}
    \caption{Example 2: AI Response Phrase Score}
    \label{fig:example2}
\end{figure}

\begin{table}[htbp]
\centering
\renewcommand{\arraystretch}{1.3} 
\caption{AI Responding Stage (Role: AI, corresponding to Example 2)}
\resizebox{\linewidth}{!}{
\begin{tabular}{l ccccccc}
\toprule
\textbf{Event} & \textbf{E 1} & \textbf{E 2} & \textbf{E 3} & \textbf{E 4} & \textbf{E 5} & \textbf{E 6} & \textbf{E 7} \\
\midrule
Time Step $t$ & $t_1$ & $t_2$ & $t_3$ & $t_4$ & $t_5$ & $t_6$ & $t_7$ \\
Time Delta $\Delta t$ & 0.5 & 0.5 & 0.5 & 0.5 & 0.5 & 1.0 & 0.5 \\
Pitch $x_{pitch}$ & 60 & 64 & 67 & 76 & 71 & 72 & 71 \\
Harmony $C_t$ & CM7 & CM7 & CM7 & CM7 & CM7 & CM7 & CM7 \\
Role $R_t$ & 0 & 0 & 0 & 0 & 0 & 0 & 0 \\
Position $P_t$ & Start & Continue & Continue & Continue & Continue & Continue & End \\
\bottomrule
\end{tabular}}
\end{table}

\begin{figure}[htbp]
    \centering
    \includegraphics[width=\linewidth]{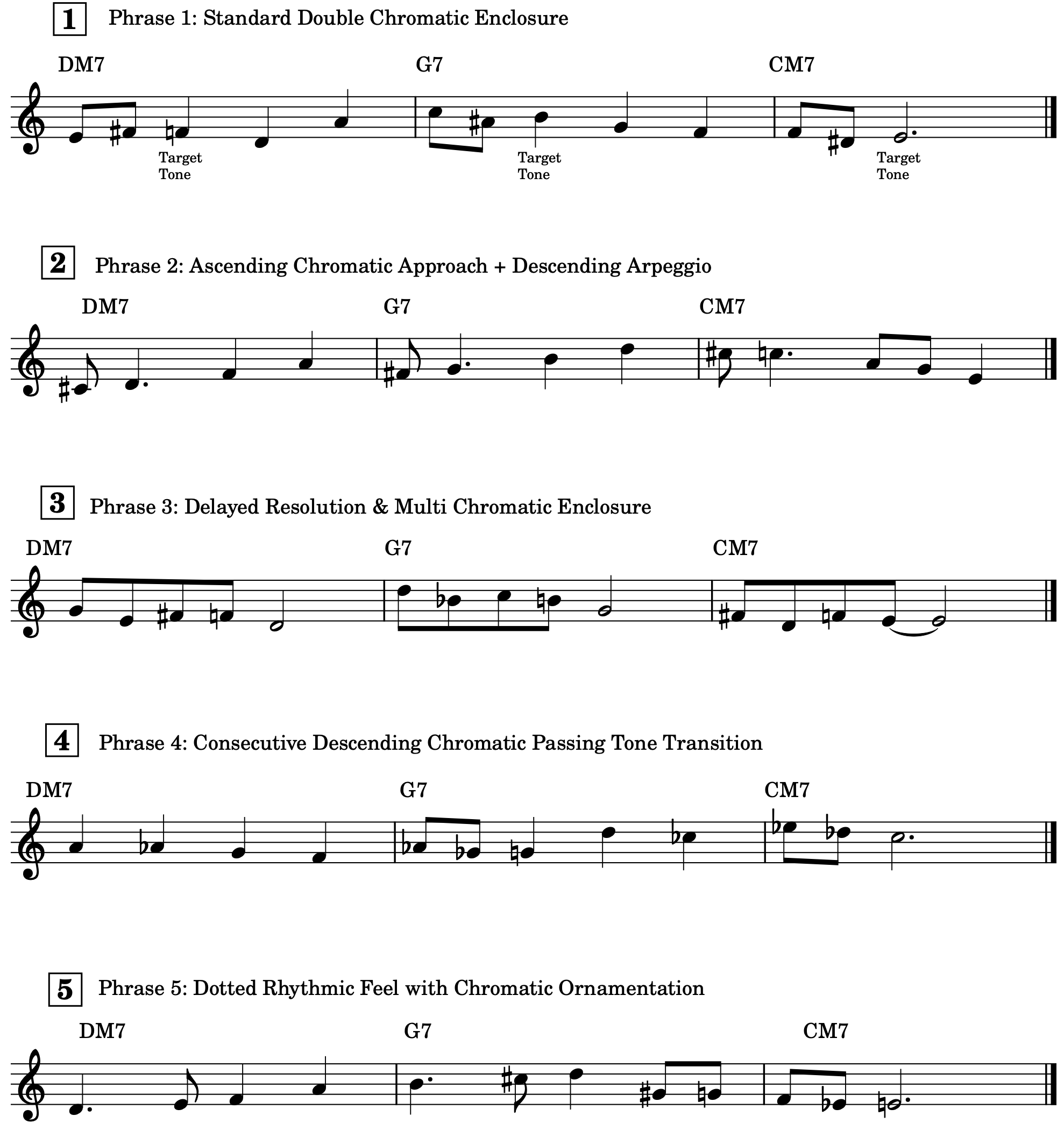}
    \caption{Example 3: Stage 2 Linear Vocabulary Expansion and Self-Motif Development}
    \label{fig:stage2_appendix}
\end{figure}

\begin{figure}[htbp]
    \centering
    \includegraphics[width=\linewidth]{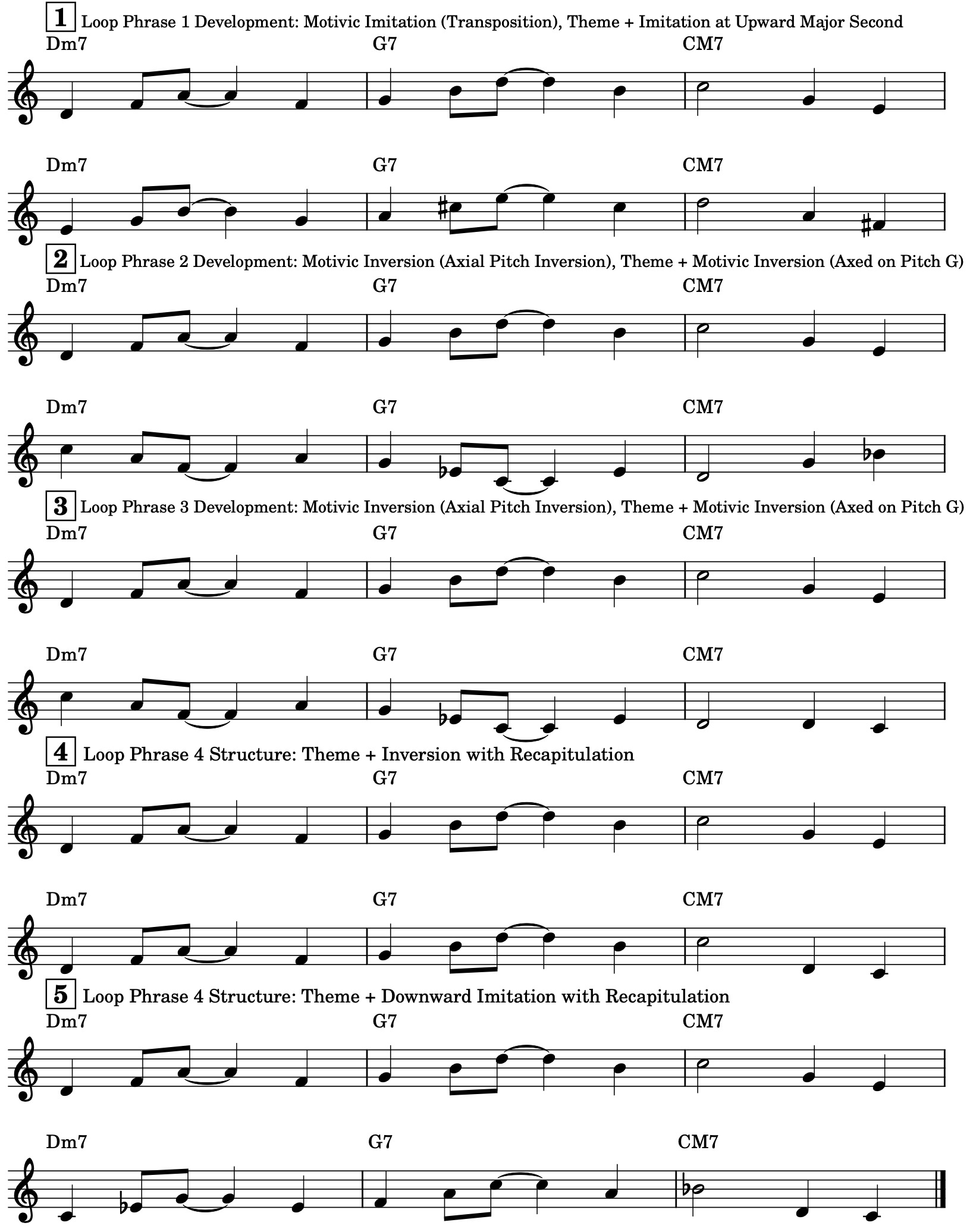}
    \caption{Example 4: Stage 3 Expert-Level Human-Computer Interactive Logic Alignment}
    \label{fig:stage3_appendix}
\end{figure}

\end{document}